\documentclass[10pt,journal]{IEEEtran}
\usepackage[cmex10]{amsmath}

\usepackage{accents}
\usepackage{color,soul}
\usepackage{verbatim}

\usepackage{amsthm}
\usepackage{amsfonts,amssymb}
\usepackage[utf8x]{inputenc}
\usepackage{flushend}
\usepackage{graphicx}
\usepackage{latexsym}
\usepackage{booktabs}
\usepackage{multirow}
\usepackage[table]{xcolor}
\usepackage{subcaption}
\usepackage{epstopdf}
\usepackage{footnote}
\usepackage{tablefootnote}
\usepackage{hyperref}
\usepackage{algpseudocode}

\usepackage{adjustbox}
\usepackage{mathtools}
\usepackage{setspace}
\usepackage{cite}
\usepackage[margin=0.6in]{geometry}
\usepackage{float}

\usepackage[ruled,vlined]{algorithm2e}
\SetKwInput{KwInput}{Input}                
\SetKwInput{KwOutput}{Output}              
\usepackage[font={footnotesize}]{caption}

\usepackage{etoolbox}

\begin{document}

\title{\huge 
Distributed and Asynchronous Operational Optimization of Networked Microgrids\\
}

\author{Nima~Nikmehr,~\IEEEmembership{Student~Member,~IEEE,} 
        Mikhail~A. Bragin,~\IEEEmembership{Member,~IEEE,}
                Peter~B. Luh,~\IEEEmembership{Life~Fellow,~IEEE,} and~Peng~Zhang,~\IEEEmembership{Senior~Member,~IEEE}
\thanks{This work was supported in part by the National Science Foundation under Grant ECCS-2018492, CNS-2006828 and OIA-2040599.}                
\thanks{N. Nikmehr and P. Zhang are with the Department of Electrical Engineering and Computer Science, Stony Brook University, Stony Brook, NY 11794, USA (e-mail: p.zhang@stonybrook.edu).}
\thanks{M. A. Bragin and  P. B. Luh are with  the  Department  of Electrical  and  Computer  Engineering,  University  of  Connecticut,  Storrs,  CT06269, USA.}
}

\maketitle

\begin{abstract}
Smart programmable microgrids (SPM) is an emerging technology for making microgrids more software-defined and less hardware-independent such that converting distributed energy resources (DERs) to networked community microgrids becomes affordable, autonomic, and secure. 
As one of the cornerstones of SPM, this paper pioneers a concept of software-defined operation optimization for networked microgrids, where operation objectives, grid connection, and DER participation will be defined by software and plug-and-play, and can be quickly reconfigured, based on the development of modularized and tightened models and a novel asynchronous price-based decomposition-and-coordination method. 
Key contributions include: (1) design the architecture of the operational optimization of networked microgrids which can be readily implemented to ensure the programmability of islanded microgrids in solving the distributed optimization models, (2) realize a novel discrete model of droop controller, and
(3) introduce a powerful distributed and asynchronous method Distributed and Asynchronous Surrogate Lagrangian Relaxation (DA-SLR) to efficiently coordinate microgrids asynchronously. Two case studies are tested to demonstrate the efficiency of developed DA-SLR, and specifically, the testing results show the superiority of DA-SLR as compared to previous methods such as ADMM.
\end{abstract}

\IEEEpeerreviewmaketitle

\normalsize

\section{Introduction}\label{Section:01}
 The smart programmable microgrids (SPMs) is the emerging phenomenon to address the issues associated with the existing drawbacks of microgrids' (MGs) structure such as the dependence on hardware, challenges in network virtualization, and vulnerability of communication signals to cyber-attacks~\cite{NSFproposal}. 
To provide flexible and easily manageable distributed and asynchronous operational optimization of islanded networked microgrids, software-defined networking (SDN) has been used. Within SDN, the network programmability is enabled through the use of logically centralized controllers \cite{6994333}. The programmability of SDN allows the network to efficiently manage the communication signals, and to enable the user access to the switches to manage the network after detecting the failures owing to the data plane and control plane separation.
Since MGs necessitate the exploitation of SDN in the communication network, the SDN realizes software and plug-and-play-based operation objectives, grid connection, and distributed energy resources (DER) participation. In~\cite{zhang2021networked}, an SDN-based MG framework is designed to manage a self-healing network and enhance the resilience of the network. To ensure resilient microgrids operations, SDN-based communication architecture is developed in \cite{zhang2019enabling} to manage the cyber-physical disturbances.

Within the SDN infrastructure, discrete operations play an important role and the control signals are sent through the switches as packets. Further 
, in programmable MGs, the discrete operation mode can also help the network operators to accurately understand the network dynamics even in presence of communication delays in real applications~\cite{9099877}. Discrete controllers possess the desired features such as simple programming, cost-efficiency, and digital and analog input and outputs~\cite{4488380}. Discrete controllers are the prevailing control mode in microgrids to appropriately manage the DERs dispatch, substantially resolved the intractable efforts in guaranteeing microgrid stability. In~\cite{7870716,7592418}, to control the frequency and voltage deviations, distributed discrete secondary control are used. 

Since the computation burden in the centralized operation of networked MGs increases with the increase of the network size, to coordinate distributed entities, distributed optimization methods have been used to improve computational performance and to resolve data privacy issues of centralized methods \cite{6980137}. Within the distributed methods, the problem is decomposed into several subproblems thereby ensuring the privacy of entities, and the avoidance of single-point failures.
The Lagrangian relaxation (LR) method is suitable for distributed coordination. Within the method, after constraints that couple distributed entities are relaxed, the relaxed problem is split into several subproblems, which are coordinated by updating Lagrangian multipliers. 
To accelerate the convergence of the LR method, Augmented Lagrangian relaxation (ALR) \cite{GEORGES1994155} has been used by penalizing the violations \cite{HADIAMINI2018137}. However, within the method, the problem is non-separable and nonlinear. To overcome the non-separability issue, the alternate direction method of multipliers (ADMM) was used \cite{5960802}. While within the ADMM subproblems are smaller in size and are easier to solve than the relaxed problem within ALR, the objective function of each subproblem includes a quadratic penalty entailing the decision variables from other subproblems which leads to communication and privacy issues.
To coordinate distributed entities without spending the time for synchronization, in \cite{8682117}, an asynchronous ADMM algorithm was used to allow the coupling variables to be updated in each subproblem without getting updates from other subsystems. However, the ADMM does not converge in the presence of binary variables.

To tackle the above issues, we contribute the following:

\begin{itemize}
    \item The architecture of software-defined networking is established to prepare the implementation and easy management of distributed operational optimization of programmable microgrids in future studies.
    \item Energy management of networked microgrids is formulated as a distributed operational optimization problem considering the operational limits and power flow constraints. The droop controllers are discretized for simple programmability and high compatibility in the SDN framework due to sending the control signals as packets.
    \item A distributed and asynchronous surrogate Lagrangian relaxation method \cite{9112667} is employed to coordinate the interconnected microgrids. Within DA-SLR, each sub-system shares data only with the coordinator without sending data to neighboring sub-systems or MGs. This sharing policy of DA-SLR preserves privacy. Also, compared to the classical distributed methods, 
    our DA-SLR method ensures the convergence in the presence of discrete variables.
\end{itemize}

\section{SDN-enabled energy management of islanded Microgrids}\label{Section:03}
\subsection{System model of MGs}\label{Subsection01}
\begin{figure}
	\centering
	\includegraphics[width=0.5\textwidth]{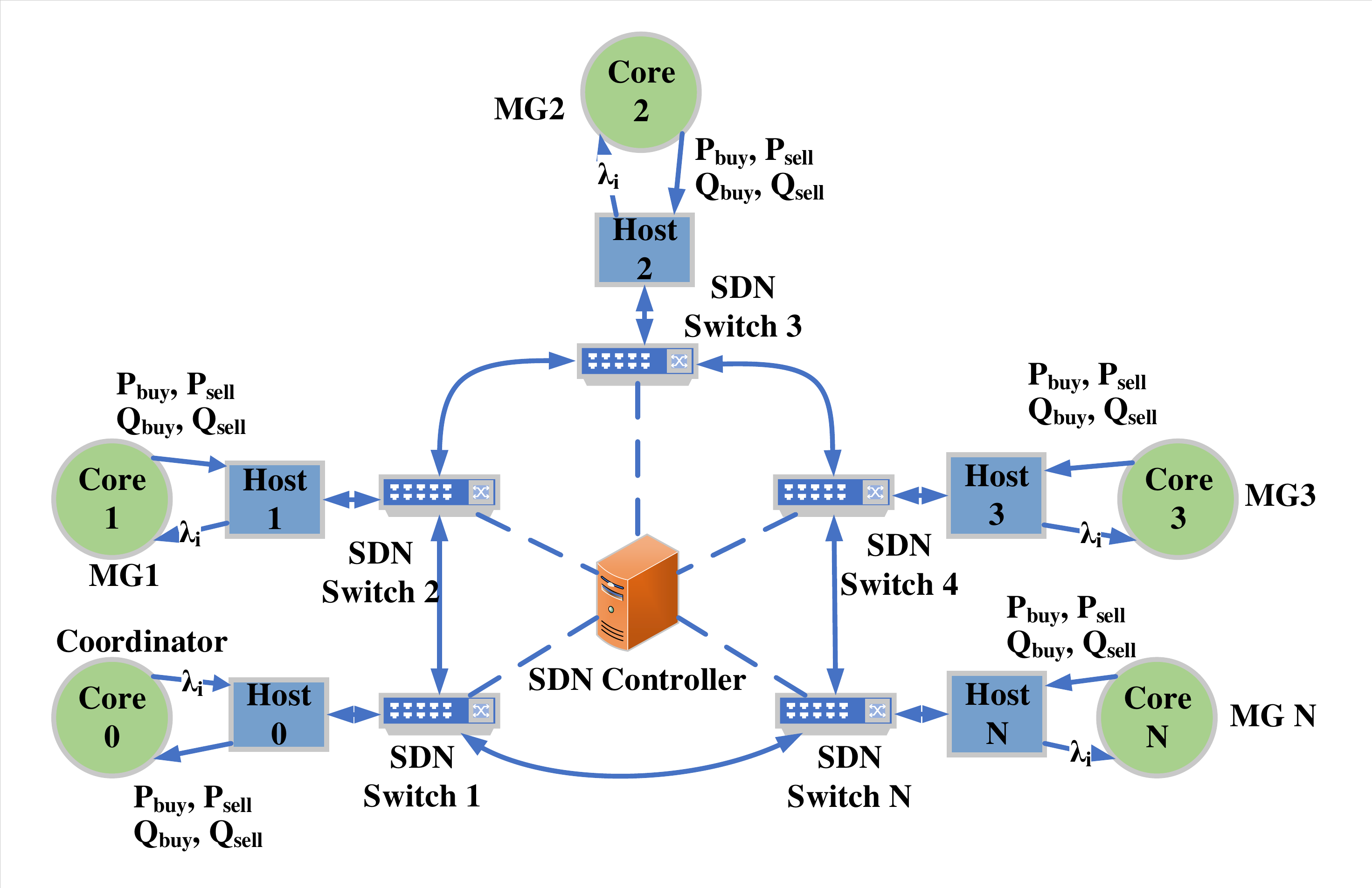}
	\caption{\footnotesize Scheme of SDN-enabled operational optimization of networked MGs}
	\label{figAdd01}\vspace{-10px}
\end{figure}

Software-defined networking takes the role of the data transfer between microgrids and the coordinator. Within the SDN architecture, switches are converted to faster and easy forwarding devices owing to the separation of data and control planes. Besides, the control technique is the centralized operating system \cite{6994333}. As shown in Fig. \ref{figAdd01}, the SDN structure includes several interconnected switches and only controllers with a wide view of the network are selected to route data transfer \cite{9007663}.

\begin{figure}
  \centering
  \includegraphics[width=0.5\textwidth]{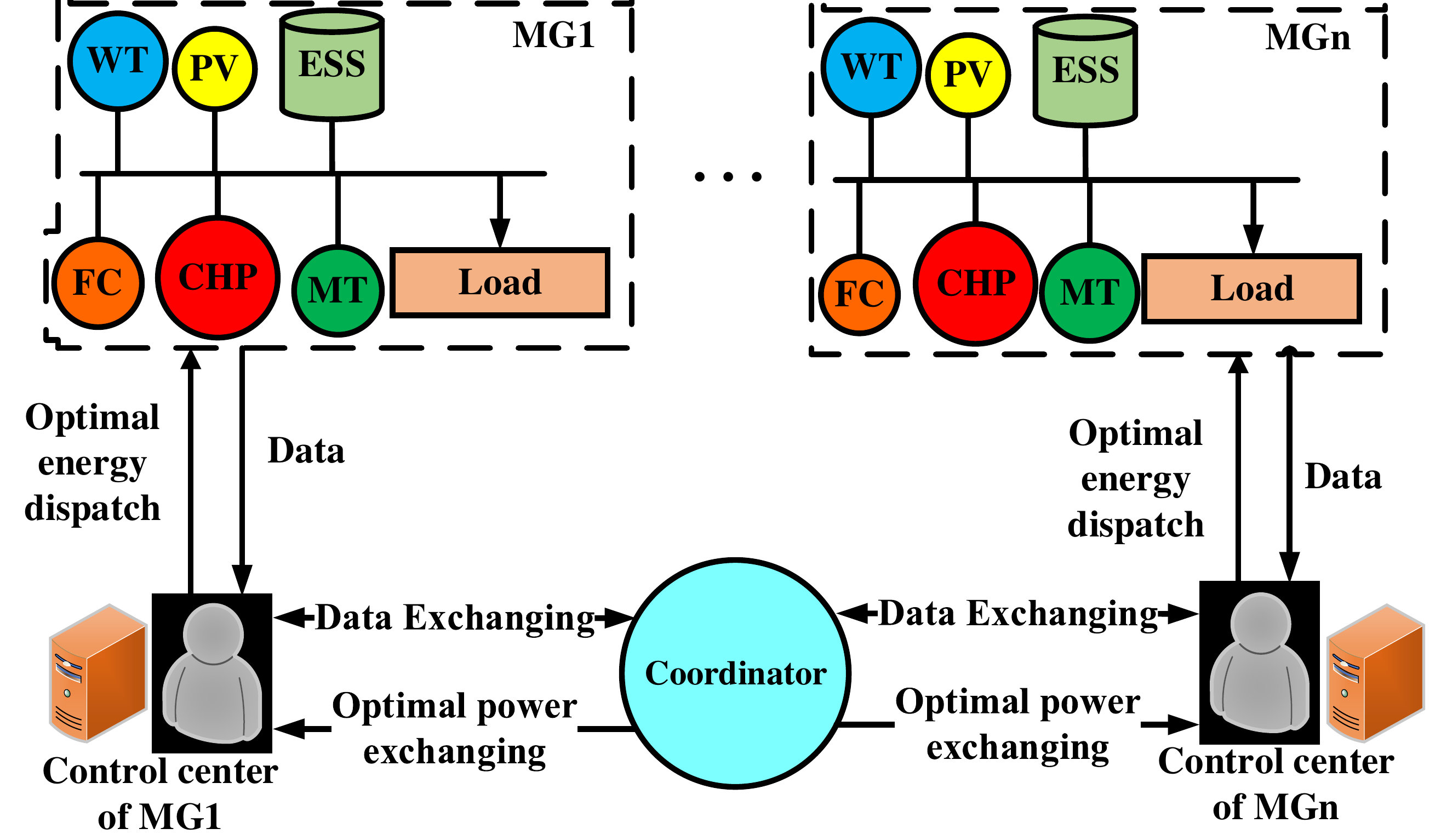}
\caption{\footnotesize Networked MG-based distribution grid }
\label{fig01}
\end{figure}

In the designed model, a networked structure of islanded microgrids is considered. Each microgrid is connected to one or a group of neighbouring MGs. Therefore, the control center of MGs should collect data from their local controllers to make an accurate decision about the power exchanging possibility with neighbouring MGs. Each microgrid consists of dispatchable and nondispatchable generators and loads. In this structure, MG control center observes the difference between generation and load, and then make a decision about power exchanging with neighbour MGs. In the presented networked MGs structure, each MG is considered as an autonomous entity. To guarantee the operational optimization of networked MGs, all MG entities should be coordinated. Thus, a distributed algorithm is employed to coordinate the MGs operation. In Fig. \ref{fig01}, a general description for optimal connection between islanded MGs is shown. According to Fig. \ref{figAdd01}, the shared data with the coordinator are the amount of purchased and sold real and reactive powers which are determined after MGs optimization. 

\subsection{Optimization model of islanded MGs}\label{Subsection02}
\textbf{Objective function.} In this subsection, the optimization model consisting of objective function and technical constrains is presented for the networked microgrids. 

The objective function is described as:
\begin{equation}
\label{eq09}
\begin{split}
OF=& \sum_{t \in T}\sum_{i\in I}[C_{g}(P^{g,MT}_{t,i})+C_{g}(P^{g,FC}_{t,i})+C_{g}(P^{g,CHP}_{t,i})]\\
& +\sum_{t \in T}\sum_{b \in B}[C_{dch}(P^{BAT,dch}_{t,b})-C_{ch}(P^{BAT,ch}_{t,b})]\\
& +\sum_{m \in M}\sum_{t \in T}[\zeta_{IL}\cdot P_{t,m}^{IL}+\zeta_{IL}\cdot Q_{t,m}^{IL}]
\end{split}
\end{equation}
where, 
\begin{subequations}
    \begin{align}
        C_{g}(P^{g,MT})&=\alpha_{MT}^{g}\cdot P^{g,MT}\label{eq01a}\\
        C_{g}(P^{g,FC})&= \alpha_{FC}^{g}\cdot P^{g,FC}\label{eq01b}\\
        C_{g}(P^{g,CHP})&= \alpha_{CHP}^{g}\cdot P^{g,CHP}\label{eq01c}
    \end{align}
\end{subequations} 
are generation costs of micro turbine (MT), fuel cell (FC) and combined heat and power (CHP), respectively, with $P^{g,MT}$, $P^{g,FC}$ and $P^{g,CHP}$ being their corresponding power generation levels and  $\alpha_{MT}^{g}$, $\alpha_{FC}^{g}$ and $\alpha_{CHP}^{g}$ being their generation prices. The costs $C_{ch}$ and $C_{dch}$ of charging and discharging level of batteries are calculated as follows:
\begin{subequations}
    \begin{align}
        C_{ch}(P^{BAT,ch})&= \alpha_{ch}^{BAT}\cdot P^{BAT,ch}\label{eq02a}\\
        C_{dch}(P^{BAT,dch})&= \alpha_{dch}^{BAT}\cdot P^{BAT,dch}\label{eq02b}
    \end{align}
\end{subequations}
where, $P^{BAT,ch}$ and $P^{BAT,dch}$ are charging and discharging power of batteries, respectively and  $\alpha_{ch}^{BAT}$ and $\alpha_{dch}^{BAT}$ are the charging and discharging prices. Decision variables $P^{IL}_{t,m}$ and $Q^{IL}_{t,m}$ are real and reactive power load shedding in microgrid $m$ at hour $t$. Also, $\zeta_{IL}$ is load shedding price and $m$ is used to describe microgrid index. In the above, sets $I$, $B$, $M$ and $T$ are used to denote sets of generation units, batteries, microgrids and time periods, respectively.

As shown in (\ref{eq09}), the first summation is the total generation cost, the second summation is the total battery charging and discharging cost and the third summation is the total real and reactive powers interruption cost. 

\textbf{Constraints.} The objective function (\ref{eq09}) is minimized subject to the following technical constraints.

\begin{itemize}
    \item \textbf{Generation Capacity.}
\end{itemize} The constraints restricting real and reactive power generation levels are defined as follows:
\begin{subequations}
\begin{equation}
u^{g}_{t,i}\cdot \underline{P}^{g}\leq P^{g}_{t,i}\leq  u^{g}_{t,i}\cdot \overline{P}^{g}, \quad \forall t\in T, i\in I\label{eq06a}
\end{equation}
\begin{equation}
u^{g}_{t,i}\cdot \underline{Q}^{g}\leq Q^{g}_{t,i}\leq  u^{g}_{t,i}\cdot \overline{Q}^{g}, \quad \forall t\in T, i\in I\label{eq06b}
\end{equation}
where, $u^{g}_{t,i}$ defines commitment status of $i^{th}$ DG at hour $t$, $P_{t,i}^{g}$ and $Q_{t,i}^{g}$ are real and reactive power generation by unit $i$ at hour $t$. $\underline{P}^{g}$, $\overline{P}^{g}$, $\underline{Q}^{g}$, and $\overline{Q}^{g}$ are minimum generated real power, maximum generated real power, minimum generated reactive power, and maximum generated reactive power, respectively.

\begin{itemize}
    \item \textbf{Charge/Discharge Power Limits.}
\end{itemize} The charging and discharging status of batteries should follow the charging limits as below:
\begin{equation}
0\leq P_{t,b}^{BAT,ch}\leq  u_{t,b}^{ch}\cdot \overline{P}^{ch}, \quad \forall t\in T, b\in B\label{eq06g}
\end{equation}
\begin{equation}
0\leq P_{t,b}^{BAT,dch}\leq  (1-u_{t,b}^{ch})\cdot \overline{P}^{dch}, \quad \forall t\in T, b\in B\label{eq06h}
\end{equation}
where, $u_{t,b}^{ch}$ is a binary variable showing the status of $b^{th}$ battery which can be either charged $(u_{t,b}^{ch} = 1)$ or discharged $(u_{t,b}^{ch} = 0)$ in each time slot. Besides, $\overline{P}^{ch}$, $\overline{P}^{dch}$ are maximum allowed amount of battery charging and discharging, respectively.

\begin{itemize}
    \item \textbf{Power Flow Limits.}
\end{itemize} Each distribution line between two nodes can distribute real and reactive power within a range between zero and $\overline{P}^{flow}$, and $\overline{Q}^{flow}$ as follows:
\begin{equation}
0\leq P_{t,n-k}^{flow}\leq  \overline{P}^{flow}, \quad \forall  t\in T, (n,k)\in N\label{eq06i}
\end{equation}
\begin{equation}
0\leq Q_{t,n-k}^{flow}\leq  \overline{Q}^{flow}, \quad \forall t\in T, (n,k)\in N\label{eq06j}
\end{equation}

\begin{itemize}
    \item \textbf{Power Interruption Constraints.}
\end{itemize} The interrupted real and reactive power levels cannot exceed the limits $\overline{P}^{IL}$ and $\overline{Q}^{IL}$:
\begin{equation}
0\leq P_{t,m}^{IL}\leq  \overline{P}^{IL}, \quad \forall t\in T, m\in M\label{eq06k}
\end{equation}
\begin{equation}
0\leq Q_{t,m}^{IL}\leq  \overline{Q}^{IL}, \quad \forall t\in T, m\in M\label{eq06l}
\end{equation}

\begin{itemize}
    \item \textbf{Voltage and Frequency Restrictions.}
\end{itemize} Voltage and frequency restrictions are described as follows:
\begin{equation}
\underline{|V_n|} \leq |V_{t,n}|\leq  \overline{|V_n|}, \quad \forall t\in T, n\in N\label{eq06m}
\end{equation}
\begin{equation}
59.5\leq f_{t}\leq 60.5, \quad \forall t\in T\label{eq06n}
\end{equation}
\end{subequations}
where, the node voltage is restricted between $\underline{|V_n|}$ and $\overline{|V_n|}$.

\begin{itemize}
    \item \textbf{Droop Control.}
\end{itemize} Droop control is deployed to enhance the ability of MGs in ensuring the real and reactive power balances. In this regard, the $f-P$ and $V-Q$ characteristics of a droop control can be described as follows \cite{4118327}:
\begin{subequations}
    \begin{align}
        f&=f_{ref}-m_{p}\cdot(P^{g}_{i}-P^{g}_{i,ref})\label{eq03a}\\
        |V|&=|V_{ref}|-m_{q}\cdot(Q^{g}_{i}-Q^{g}_{i,ref})\label{eq03b}
    \end{align}
\end{subequations}
where, $f$ and $f_{ref}$ are fluctuated frequency and reference frequency, which is set to 60 Hz. Similarly, $|V|$ and $|V_{ref}|$ are bus voltage magnitude and nominal voltage magnitude, respectively. Also, $P^{g}_i$ and $Q^{g}_i$ are real and reactive power of DGs. Droop control coefficients for frequency and voltage based controller are $m_{p}$ and $m_{q}$, respectively. The limits dealing with $m_{p}$ and $m_{q}$ are described as following:
\begin{subequations}
    \begin{align}
        \underline{m}_{p}\leq m_{p} \leq \overline{m}_{p} \label{eq003a}\\
        \underline{m}_{q}\leq m_{q} \leq \overline{m}_{q} \label{eq003b}
    \end{align}
\end{subequations}
where, $\underline{m}_{p}$ and $\overline{m}_{p}$ are minimum and maximum values of frequency-based droop controller respectively. Also, minimum and maximum values of voltage-based droop controller are described by $\underline{m}_{q}$ and $\overline{m}_{q}$ respectively.

\begin{itemize}
    \item \textbf{Power flow.}
 
\end{itemize} For the power flow study, the linearized DistFlow model is deployed. The model has been utilized and justified in distribution systems and microgrids studies \cite{6152194}-\cite{6497085}. An exact approximation of AC power flow is considered in linearized DistFlow model \cite{6895183}:
\begin{subequations}
    \begin{equation}
    \begin{split}
        P_{t,n-k}^{flow}=\sum_{(\kappa:(k,\kappa))\in N}P_{t,k-\kappa}^{flow}-(P_{t,k}^{g}-P_{t,k}^{L}),\\
        \quad\forall t\in T,\ n,k,\kappa\in N\label{eq04a}
        \end{split}
    \end{equation}
    \begin{equation}
    \begin{split}
        Q_{t,n-k}^{flow}=\sum_{(\kappa:(k,\kappa))\in N}Q_{t,k-\kappa}^{flow}-(Q_{t,k}^{g}-Q_{t,k}^{L}), \\
        \quad\forall t\in T,\ n,k,\kappa\in N\label{eq04b}
        \end{split}
    \end{equation}
    \begin{equation}
    \begin{split}
        |V_{n,t}|=|V_{k,t}|-\frac{r_{n,k}\cdot P_{t,n-k}^{flow}+\chi_{n,k}\cdot Q_{t,n-k}^{flow}}{|V_{0}|} \\
\quad\forall t\in T,\ n,k\in N\label{eq04c}
        \end{split}
    \end{equation}
\end{subequations}
where, the bus index is described by $n$ and $k$ and the set $N$ denotes a set of nodes. The real and reactive power flow between sending node $n$ and receiving node $k$ at hour $t$ are defined by $P_{t,n-k}^{flow}$ and $Q_{t,n-k}^{flow}$, respectively. Also, real and reactive loads are denoted by $P_{t,k}^{L}$ and $Q_{t,k}^{L}$, respectively, and $|V_{n,t}|$ and $|V_{0}|$ are voltage magnitudes at bus $n$ and the point of common coupling in an MG, respectively. The resistance and reactance between lines $n$ and $k$ are determined by $r_{n,k}$ and $\chi_{n,k}$. 

\begin{itemize}
 
    \item \textbf{Real power balance.}
\end{itemize} The real power balance ensures the balance between the real power consumption and real power generation and exchanging real power through the distribution lines which is formulated as follows:
\begin{equation}
\label{eq24}
\begin{split}
&\sum_{i\in I}(s_{n}^{g}\cdot P_{t,i}^{g})+s_{n}^{WT}\cdot P_{t}^{g,WT}+s_{n}^{PV}\cdot P_{t}^{g,PV}\\
&+\frac{1}{m_{p}}(f_{ref,m}-f_{t,m})+s_{n}^{BAT}(P_{t}^{BAT,dch}-P_{t}^{BAT,ch})\\
&+P_{n,t}^{IL}+(P_{t,k-n}^{flow}-\sum_{(k:(n,k))\in N}P_{t,n-k}^{flow})+\\
&+(\sum_{m \in M}(s_{n-m}^{Tran}\cdot P_{t,n-m}^{buy}-s_{n-m}^{Tran}\cdot P_{t,n-m}^{sell}))=P_{t,n}^{L},\\
&\quad\forall t\in T, (n,k)\in N, m\in M
\end{split}
\end{equation}
where, $P_{t}^{g,WT}$ and $P_{t}^{g,PV}$ are real power generation by WT and PV panel. Also, $s_{n}^{g}$, $s_{n}^{WT}$, $s_{n}^{PV}$, $s_{n}^{BAT}$ and $s_{n-m}^{Tran}$ are connection indicator of a generator, WT, PV, battery, and transacted power of node $n$ to a node in $m^{th}$ microgrid, respectively.
 
\begin{itemize}
 
    \item \textbf{Reactive power balance.}
\end{itemize} The reactive power balance denotes the balance between the reactive power generation and loads considering the reactive power flow and voltage-based droop control impact as follows:

\begin{equation}
\label{eq25}
\begin{split}
&\sum_{i\in I}(s_{n}^{g}\cdot Q_{t,i}^{g})+s_{n}^{WT}\cdot Q_{t}^{g,WT}+s_{n}^{PV}\cdot Q_{t}^{g,PV}\\
&+\frac{1}{m_{q}}(|V|_{n}^{ref}-|V_n,t|)+(Q_{t,k-n}^{flow}-\sum_{(k:(n,k))\in N}Q_{t,n-k}^{flow})\\
&+(\sum_{m \in M}(s_{n-m}^{Tran}\cdot Q_{t,n-m}^{buy}-s_{n-m}^{Tran}\cdot Q_{t,n-m}^{sell}))+Q_{n,t}^{IL}\\
&=Q_{t,n}^{L}, \quad\forall t\in T, (n,k)\in N, m\in M
\end{split}
\end{equation}
where, $Q_{t}^{g,WT}$ and $Q_{t}^{g,PV}$ are reactive power generation by WT and PV panel. 

\begin{itemize}
    \item \textbf{Interface Power Exchange Limits.}
\end{itemize} The amount of exchanging power by each MG is limited as follows:
\begin{equation}
0\leq P_{t,m-w}^{buy}\leq  u_{t,m-w}^{buy}\cdot \overline{P}^{buy}, \forall t\in T, \{m,w\}\in M\label{eq06c}
\end{equation}
\begin{equation}
0\leq P_{t,m-w}^{sell} \leq  (1-u_{t,m-w}^{buy})\cdot \overline{P}^{sell}, \forall t\in T, \{m,w\}\in M\label{eq06d}
\end{equation}
\begin{equation}
0\leq Q_{t,m-w}^{buy}\leq  u_{t,m-w}^{buy}\cdot \overline{Q}^{buy}, \forall t\in T, \{m,w\}\in M\label{eq06e}
\end{equation}
\begin{equation} 
0\leq Q_{t,m-w}^{sell}\leq  (1-u_{t,m-w}^{buy})\cdot \overline{Q}^{sell}, \forall t\in T, \{m,w\}\in M\label{eq06f}
\end{equation}
where, $P_{t,m-w}^{buy}$ and $P_{t,m-w}^{sell}$ are purchased/sold power levels by $m^{th}$ microgrid at hour $t$ from/to $w^{th}$ microgrid. Similarly, $Q_{t,m-w}^{buy}$ and $Q_{t,m-w}^{sell}$ are purchased/sold reactive power levels, respectively. Also, $u_{t,m-w}^{buy}$ is the binary variables denoting the status of purchasing power by $m^{th}$ MG from $w^{th}$ MG.
The maximum amount of purchased and sold real and reactive powers are illustrated by $\overline{P}^{buy}$, $\overline{P}^{sell}$, $\overline{Q}^{buy}$ and $\overline{Q}^{sell}$, respectively.

\begin{itemize}
    \item \textbf{Interface Power Flow Constraints.}
\end{itemize} The following interface power flow constraints are considered to ensure that the power bought by microgrid $m$ from microgrid $w$ equals to the power sold by microgrid $w$ to microgrid $m$:
\begin{equation}
P_{t,m-w}^{buy}=P_{t,w-m}^{sell}, \quad \forall t\in T, \{m,w\}\in M\label{eq06newg}
\end{equation}
\begin{equation}
Q_{t,m-w}^{buy}=Q_{t,w-m}^{sell}, \quad \forall t\in T, \{m,w\}\in M\label{eq06newh}
\end{equation}

These constraints are coupling with respect to the microgrids. 

\subsection{Problem linearization}\label{Subsection03}
The optimization solvers such as CPLEX and Gurobi cannot solve nonlinear problems due to cross products of variables. In (\ref{eq24}), frequency $f_{t,m}$ is a continuous variable while $\frac{1}{m_p}$ is discrete. Similarly, in (\ref{eq25}), voltage magnitude $|V_{n,t}|$ and $\frac{1}{m_q}$ are continuous and discrete variables, respectively. Therefore, the droop control terms make the problem constraints nonlinear, which results in a nonlinear optimization problem \cite{Sherali2013}. Nevertheless, the presented terms are linearized as described below. 

\textbf{Generic Linearization Procedure.} The linerization procedure will be explained by using an expression $z=A\cdot d$, where $A$ is a continuous variable and $d$ is a binary variable. If $A$ has bounds $[\underaccent{\bar}{A},\bar{A}]$,  then the exact form of the linearized inequalities are as following for expression $d$ \cite{Sherali2013}:
\begin{subequations}
\begin{equation}
min\{0,\underaccent{\bar}{A}\}\leq z\leq \bar{A}\label{eq09a}
\end{equation}
\begin{equation}
\underaccent{\bar}{A}\cdot d\leq z\leq \bar{A}\cdot d\label{eq09b}
\end{equation}
\begin{equation}
A-(1-d)\cdot \bar{A}\leq z\leq A-(1-d)\cdot \underaccent{\bar}{A}\label{eq09c}
\end{equation}
\begin{equation}
z\leq A+(1-d)\cdot \bar{A}\label{eq09d}
\end{equation}
\end{subequations}

Therefore, products of binary variable and continuous variables is linearized following the procedure (\ref{eq09a})-(\ref{eq09d}) through the introduction of new continuous variables ($z$). However, in the designed optimization model, $d$ is a discrete variable (integer non-negative variable). If $d_{j}$ is bounded by positive integer $D$, so that $d_{j}\in \{0,1,...,D\}$, we can introduce binary variables $w_{0},w_{1},...,w_{D}$ and add the following constraints:
\begin{subequations}
\begin{equation}
\sum_{l=0}^{D}w_{lj}=1, \quad \forall j\in J\label{eq10a}
\end{equation}
\begin{equation}
d_{j}=\sum_{l=0}^{D}l\cdot w_{lj}, \quad \forall j\in J\label{eq10b}
\end{equation}

Therefore:
\begin{equation}
z_{j}=\sum_{l=0}^{D}l\cdot w_{lj}\cdot A, \quad \forall j\in J\label{eq10c}
\end{equation}
\end{subequations}


Now, we can linearize each of the products of $w_{lj} \cdot A$ based on inequalities (\ref{eq09a})-(\ref{eq09d}). In $f-P$ droop control, $m_{p}$ is a discrete value, while frequency ($f$) is a continuous variable. Since $m_{p}$ is discrete, therefore, $k_{p}=\frac{1}{m_{p}}$ is also discrete. Therefore, 
\begin{equation}
\label{eq33}
w_{p,j}=k_{p,j}\cdot f_{m}, \quad \forall j\in J
\end{equation}

Following (\ref{eq09a})-(\ref{eq09d}), to resolve the non-linearity difficulty, we replace (\ref{eq33}) by the following constraints:
\begin{equation}
\label{eq34}
z_{p,j}=\sum_{l=0}^{D}l\cdot w_{p,lj}\cdot f_{m}, \quad \forall j\in J
\end{equation}


Similarly, by introducing $k_{q}=\frac{1}{m_{q}}$, the nonlinear terms  
\begin{equation}
\label{eq35}
w_{q,j}=k_{q,j}\cdot |V_{n}|, \quad \forall j\in J
\end{equation}
are linearized as follows:
\begin{equation}
\label{eq36}
z_{q,j}=\sum_{l=0}^{D}l\cdot w_{q,lj}\cdot |V_{n}|, \quad \forall j\in J
\end{equation}
\vspace{-20pt}
\section{Distributed and Asynchronous Surrogate Lagrangian Relaxation Model for Networked MGs}\label{Section:04}
\subsection{Distributed model of Surrogate Lagrangian Relaxation}\label{Subsection0401}
In the networked microgrids structure, assume there are $m$ MGs conndected to several neighbouring MGs. To coordinate the MGs, Lagrangian multipliers are introduced first to relax constraints (\ref{eq06newg}) and (\ref{eq06newh}) that couple MGs. The Lagrangian function then becomes:
\begin{equation}
\label{eq37}\footnotesize
\begin{split}
L(P,Q,\lambda)=& \sum_{t \in T}\sum_{i\in I}[C_{g}(P^{g,MT}_{t,i})+C_{g}(P^{g,FC}_{t,i})+C_{g}(P^{g,CHP}_{t,i})]\\
& +\sum_{t \in T}\sum_{b \in B}[C_{dch}(P^{BAT,dch}_{t,b})-C_{ch}(P^{BAT,ch}_{t,b})]\\
& +\sum_{m \in M}\sum_{t \in T}[\zeta_{IL}\cdot P_{t,m}^{IL}+\zeta_{IL}\cdot Q_{t,m}^{IL}]\\
& +\sum_{t \in T, \{m,w\}\in M}\lambda_{p,t,m-w}\cdot (P_{t,m-w}^{buy}-P_{t,w-m}^{sell})\\
& +\sum_{t \in T, \{m,w\}\in M}\lambda_{q,t,m-w}\cdot (Q_{t,m-w}^{buy}-Q_{t,w-m}^{sell}),\\
& \quad \forall t\in T, i\in I, b\in B, 
\end{split}
\end{equation}
where, $\lambda_{p,t,m-w}$ and $\lambda_{q,t,m-w}$ are Lagrange multipliers associated with real and reactive power levels.

The relaxed problem is then decomposed into individual subproblems.  The formulation for ``buying" MG $m$ is formulated as:

\begin{equation}
\label{eq38}
\begin{split}
\underset{MG_{m}}{min}\{& \sum_{i\in I}[C_{g}(P^{g,MT}_{t,m,i})+C_{g}(P^{g,FC}_{t,m,i})+C_{g}(P^{g,CHP}_{t,m,i})]\\
& +\sum_{b \in B}[C_{dch}(P^{BAT,dch}_{t,m,b})-C_{ch}(P^{BAT,ch}_{t,m,b})]\\
& +[\zeta_{IL}\cdot P_{(t,m)}^{IL}+\zeta_{IL}\cdot Q_{t,m}^{IL}]
\\ & + \sum_{t \in T, \{m,w\}\in M}\lambda_{p,t,m-w}\cdot (P_{t,m-w}^{buy})\\
& +\sum_{t \in T, \{m,w\}\in M}\lambda_{q,t,m-w}\cdot (Q_{t,m-w}^{buy}), \\ & \forall i\in I, b\in B, m \in M
\end{split}
\end{equation}
Note that decision variables $P_{t,w-m}^{sell}$ and $Q_{t,w-m}^{sell}$ within \eqref{eq37} belong to subproblem $w$ and are thus not included in \eqref{eq38}. While subproblem formulation $m$ includes variables $P_{t,m-w}^{buy}$ and $Q_{t,m-w}^{buy}$ indicating power bought, the microgrid can in fact sell power, in which case the corresponding optimized values will be negative. The formulation for ``selling" MG $w$ is formulated using the same logic with the exception that variables $P_{t,w-m}^{sell}$ and $Q_{t,w-m}^{sell}$ will appear within the objective function with the negative sign. 

The MG subproblems are coordinated by Lagrangian multipliers, which are updated based on violation of relaxed constraints as:
\begin{subequations}
    \begin{align}
        & \lambda_{p,t,m-w}^{r+1}=\lambda_{p,t,m-w}^{r}+e^{r}\cdot \big(P_{t,m-w}^{buy,r}-P_{t,w-m}^{sell,r}\big), \label{eq39a} \\
        & \lambda_{q,t,m-w}^{r+1}=\lambda_{q,t,m-w}^{r}+e^{r} \cdot \big(Q_{t,m-w}^{buy,r}-Q_{t,w-m}^{sell,r}\big), \label{eq39b} 
    \end{align}
\end{subequations}
where, $r$ is the coordinator iteration number, $P_{t,m-w}^{buy,r}$ is the most recent value available of active power at iteration $r$ that MG $m$ has the intention to buy and $e^r$ is the stepsize. In \cite{9112667}, the contraction mapping concept is employed to derive $e^r$ in the following way:
\begin{equation}
\label{eq40}
\begin{split}
& e^{r}=\gamma^{r}\frac{e^{r-1}||g(P^{r-1},Q^{r-1})||}{||g(P^{r},Q^{r})||}, 0 < \gamma^{r} < 1, r=1,2,...
\end{split}
\end{equation}
where, $g(P^{r},Q^{r})$ is the vector of constraint violations, $\gamma^{r}$ is stepsize-setting parameter which can  be  updated in each iteration. 
An exact formula 
to determine $\gamma^r$ can be found in~\cite{9112667}.

As mentioned in Section \ref{Section:03} and as shown in \eqref{eq38}, MG subproblems do not involve decision variables from other subproblems, thus, unlike ADMM, the DA-SLR method does not require data exchange among MGs thereby preserving privacy. 

\subsection{SDN-enabled DA-SLR solution steps}\label{Subsection0402}

According to Fig. \ref{figAdd01}, each core is devoted to a microgrid and the multipliers are asynchronously updated using plug and play property of DA-SLR method without waiting for all subproblem solutions to arrive at the coordinator. MGs share their purchased and sold power levels with the coordinator, the coordinator updates the multipliers and then broadcasts the multipliers to all MGs. 

\textbf{Feasible Cost Search.} Feasible solutions are searched by using heuristics. Heuristics are operationalized by solving the original problem \eqref{eq09}-\eqref{eq06newh} with decision variables fixed at the most recent values obtained by solving MG subproblems.  
If the feasible solution to the original problem is not found, the multipliers are updated for several iterations before feasible solutions are searched again. 

\textbf{Dual value of DA-SLR.} Dual values provide the lower bound for the feasible cost to quantify its quality, and the dual values are obtained by minimizing each MG subproblem by using the most recent available values of multipliers.

\textbf{Stopping criteria.} The stopping criteria is the difference  between  the feasible cost and the dual value:
\begin{equation}
\label{eq42}
\begin{split}
& OG=\frac{cost^{feasible}-cost^{dual}}{cost^{feasible}}
\end{split}
\end{equation}

The steps of distributed and asynchronous of surrogate Lagrangian relaxation is described as Algorithm \ref{al02}.


\begin{algorithm}[]
\small
\SetAlgoLined
 \KwResult{Exchanging powers ($P^{buy}_{m-w}$/$P^{sell}_{m-w}$,$Q^{buy}_{m-w}$/$Q^{sell}_{m-w}$) between microgrids}
 \textbf{1} initialization\; $M$ microgrids' exchanged powers $P^{0,buy}_{m-w}$/$P^{0,sell}_{m-w}$, $Q^{0,buy}_{m-w}$/$Q^{0,sell}_{m-w}$, Lagrange multipliers $\lambda^{0}$, updating stepsize $e^{0}$  \\
 \textbf{2} iteration $r\gets 0$ \\
 \textbf{3} Coordinator receives MGs subproblem solutions and updates multipliers per \eqref{eq39a}-\eqref{eq39b} without waiting for all solutions to arrive. Coordinator then broadcasts multipliers to all MGs\\
 \textbf{4} Each idle MG starts solving its subproblem by using the latest available multipliers and sends its solution to the coordinator\\
 \textbf{5} \eIf{criteria to search to feasible solution are satisfied}{
  search for a feasible solution \\
  \textbf{6} \eIf{the feasible solution is obtained}{
  \textbf{7} obtain the dual value and calculate the gap (\ref{eq42})
  }{
  go to step \textbf{3};
  } 
  }{
  go to step \textbf{3};
  } 
  \textbf{8} \eIf{the stopping criteria is satisfied }{
  the optimization process is finished
  }{
  go to step \textbf{3};
  } 
 \caption{Execution of DA-SLR on operational optimization of networked MGs}
 \label{al02}
\end{algorithm}
\section{Simulation results}\label{Section:05}
In this section, the efficacy and efficiency of the DA-SLR based operational optimization method are tested and validated using two case studies. In case study 1, a four-MG networked microgrid system based on a modified IEEE 33-bus distribution network~\cite{25627} is considered. In the second case study, a nine-MG networked microgrid system based on a modified IEEE 123-bus system~\cite{119237}  is tested. 
In both studies, the range of frequency-based droop coefficients $m_p$ is assumed to fall in $[0.02, 0.2]$ with a discrete step of $m_p$ considered as 0.0018. Similarly, 
the range of voltage-based droop coefficients $m_q$ is set as $[0.05, 0.5]$ with a discrete step of 0.0045.

\subsection{Test results on a 33-bus networked microgrid system}\label{05:01}
\begin{figure}[ht]
	\centering
	\includegraphics[width=0.5\textwidth]{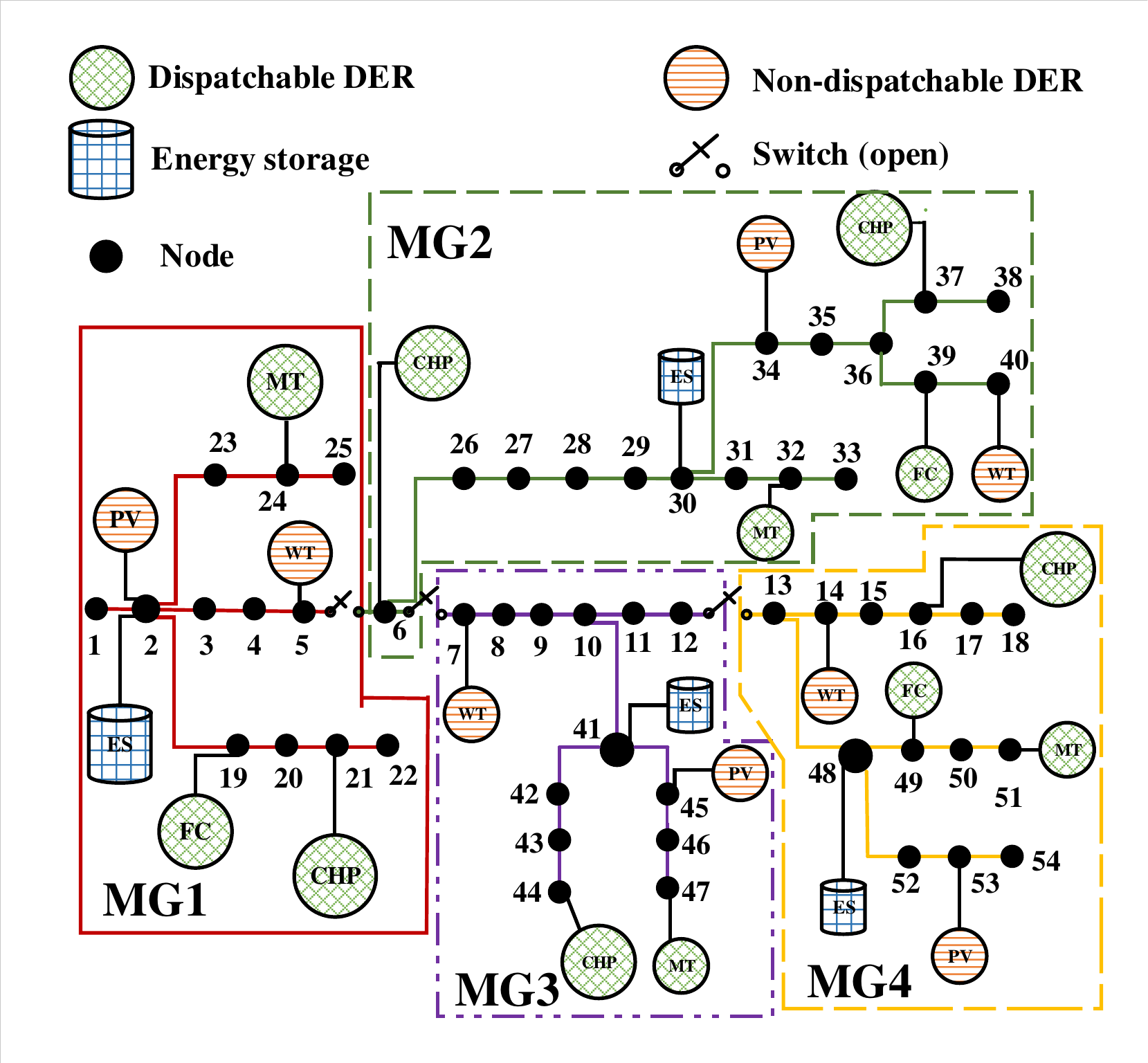}
	\caption{\footnotesize Modified IEEE 33-bus distribution system with four MGs}
	\label{fig04}\vspace{-10px}
\end{figure}

In Fig. \ref{fig04}, the single line diagram of the 33-bus networked microgrids system 
with four MGs is shown, and in Table \ref{table02}, DERs information is described.
The power base of the system is set to be 10 MVA. The line resistance and reactance of the rest of the network for added buses are set to be 0.006 and 0.01 p.u., respectively.


{\rowcolors{2}{gray!10}{lightgray!2}
\begin{table}[h]
    \centering
    \caption{DERs information in networked islanded MGs structure}
    \begin{tabular}{|p{1.0cm}||p{2.1cm}||p{1.8cm}||p{2.1cm}|}
    \rowcolor{gray!60} DER type & \multicolumn{1}{|c|}{Bus No.} & Max. real power (kW) & Max. reactive power (kVAR)  \\
    \hline\hline
    PV  &  $2,34,45,53$ & \multicolumn{1}{|c|}{$200$}  &  \multicolumn{1}{|c|}{$80$}\\
    WT  &  $5,7,14,40$ & \multicolumn{1}{|c|}{$150$}  &  \multicolumn{1}{|c|}{$60$}\\
    MT  &  $24,32,47,51$ & \multicolumn{1}{|c|}{$400$}  &  \multicolumn{1}{|c|}{$200$}\\
    FC  &  $19,39,49$ & \multicolumn{1}{|c|}{$300$}  &  \multicolumn{1}{|c|}{$-$}\\
    CHP  &  $6,16,21,37,44$ & \multicolumn{1}{|c|}{$400$} & \multicolumn{1}{|c|}{$300$}\\
    \hline
    \end{tabular}
    \label{table02}
\end{table}
}

\begin{figure}[H]
	\centering
	\includegraphics[width=0.5\textwidth]{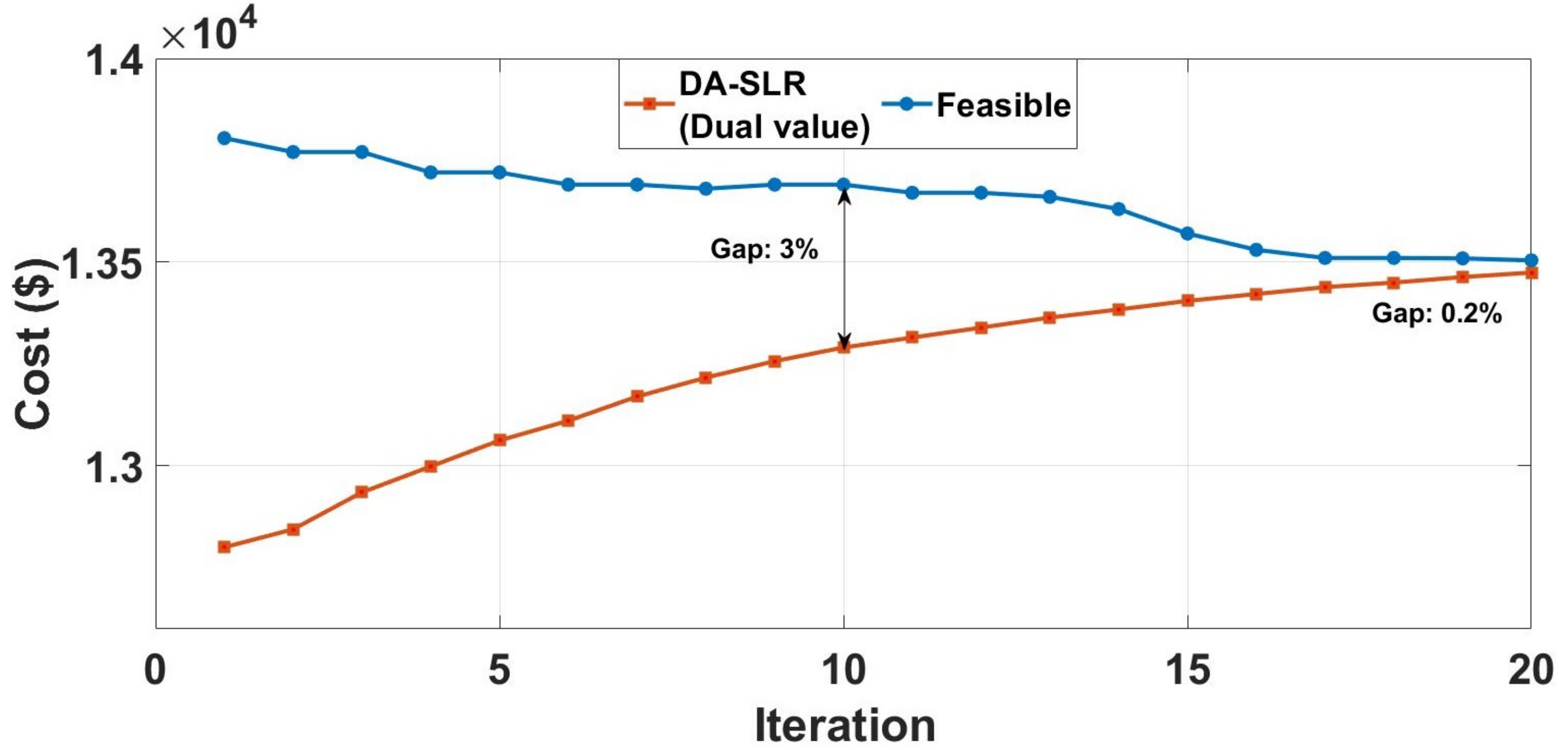}
	\caption{\footnotesize The DA-SLR method in tracking the feasible solution in modified IEEE 33-bus system 
	}
	\label{fig06}\vspace{-10px}
\end{figure}

As demonstrated in Fig. \ref{fig06}, the operation cost of the system obtained by the new method decreases fast thereby reaching the gap of less than 0.2\% within 20 iterations. 

Performance of the new method is compared against that of ADMM in Table \ref{table03}. 
{\rowcolors{2}{gray!10}{lightgray!2}
\begin{table}[h]
    \centering
    \caption{Comparison of feasible costs using DA-SLR and ADMM}
    \begin{tabular}{|p{1.4cm}||p{0.9cm}||p{0.9cm}||p{0.9cm}||p{0.9cm}||p{1.0cm}|}
    \rowcolor{gray!60} Method & \multicolumn{5}{|c|}{Objective function ($\$$)} \\
     & MG1 & MG2 & MG3 & MG4 & Total \\
    \hline\hline
    DA-SLR  &  $2119$ & $5231$  &  $3150$ &  $3062$ &  $13562$\\
    ADMM  &  $4621$ & $5759$  &  $16680$ &  $4742$ &  $31802$\\
    \hline
    \end{tabular}
    \label{table03}
\end{table}
}

The DA-SLR and ADMM methods take 16 and 102 seconds per iteration to solve MGs' subproblem, respectively. This comparison clearly shows the advantage of DA-SLR as compared to the ADMM. As reviewed in Introduction, ADMM diverges in the presence of discrete variables; and as demonstrated in Table \ref{table03}, DA-SLR obtains the total feasible cost of \$13,562, which is \$18,240 less (or 57.3\%) than that obtained by using ADMM after 20 iterations. In this case study, each iteration includes a total of four subproblems, and after finishing the optimization process of those subproblems, the next iteration starts.

\begin{figure}[H]
	\centering
	\includegraphics[width=0.5\textwidth]{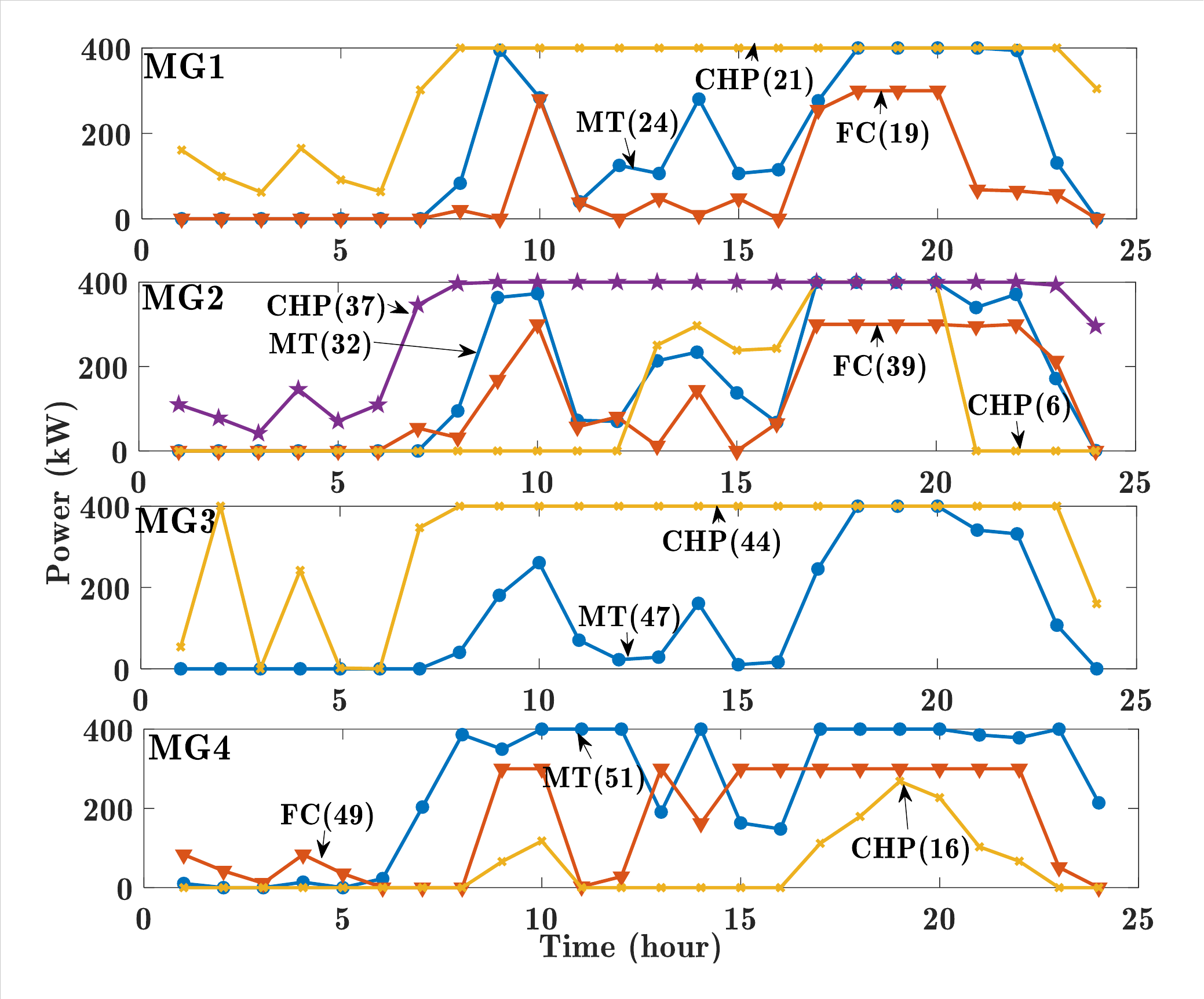}
	\caption{\footnotesize Generated real power by DERs in each microgrid obtained by DA-SLR method}
	\label{fig08}\vspace{-10px}
\end{figure}

\begin{figure}[H]
	\centering
	\includegraphics[width=0.5\textwidth]{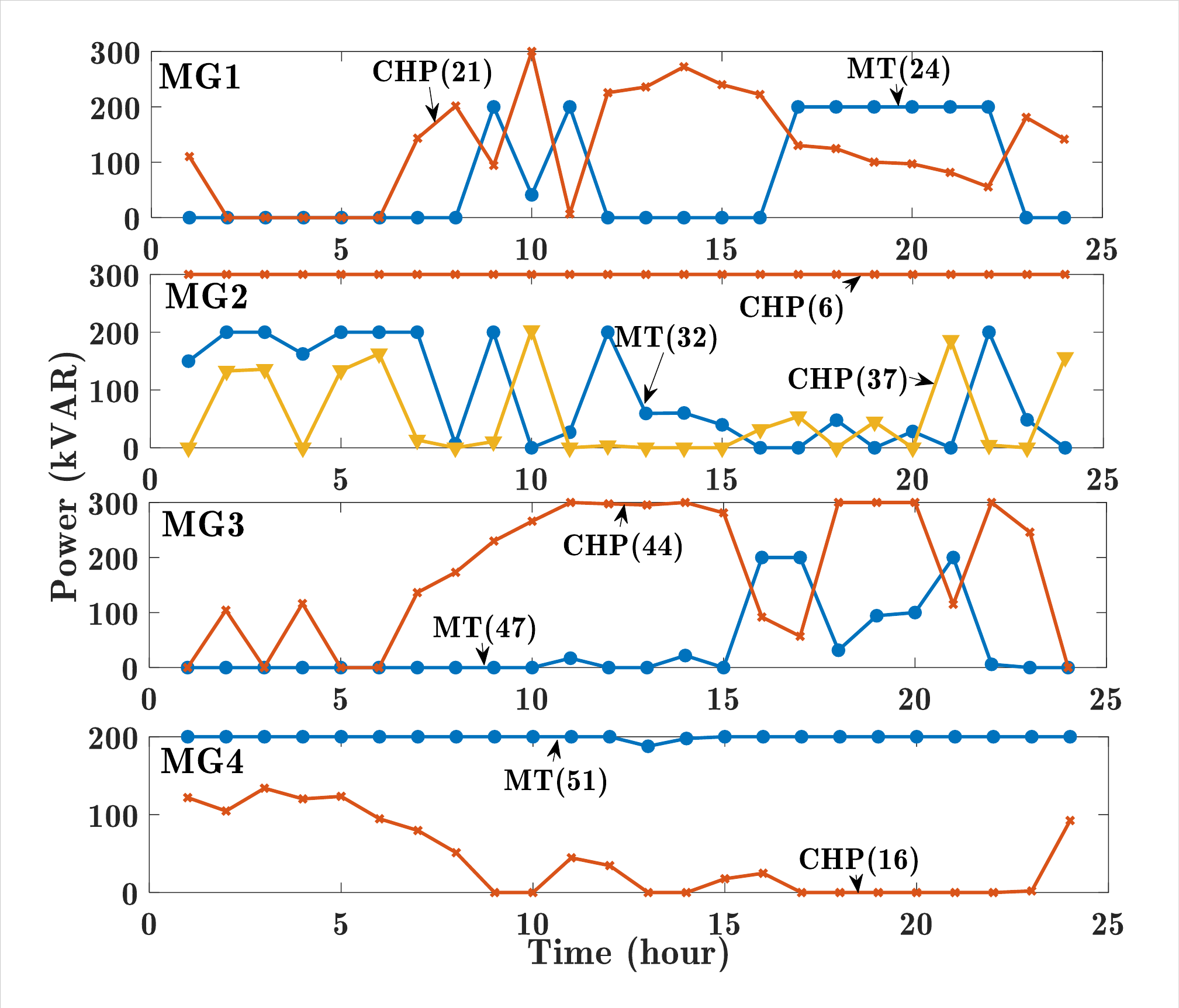}
	\caption{\footnotesize Generated reactive power by DERs in each microgrid obtained by DA-SLR method}
	\label{fig09} 
\end{figure}

The contribution of active and reactive power levels within each DER obtained by DA-SLR are shown in Fig. \ref{fig08} and Fig. \ref{fig09}, respectively.

\begin{figure}[H]
	\centering
	\includegraphics[width=0.5\textwidth]{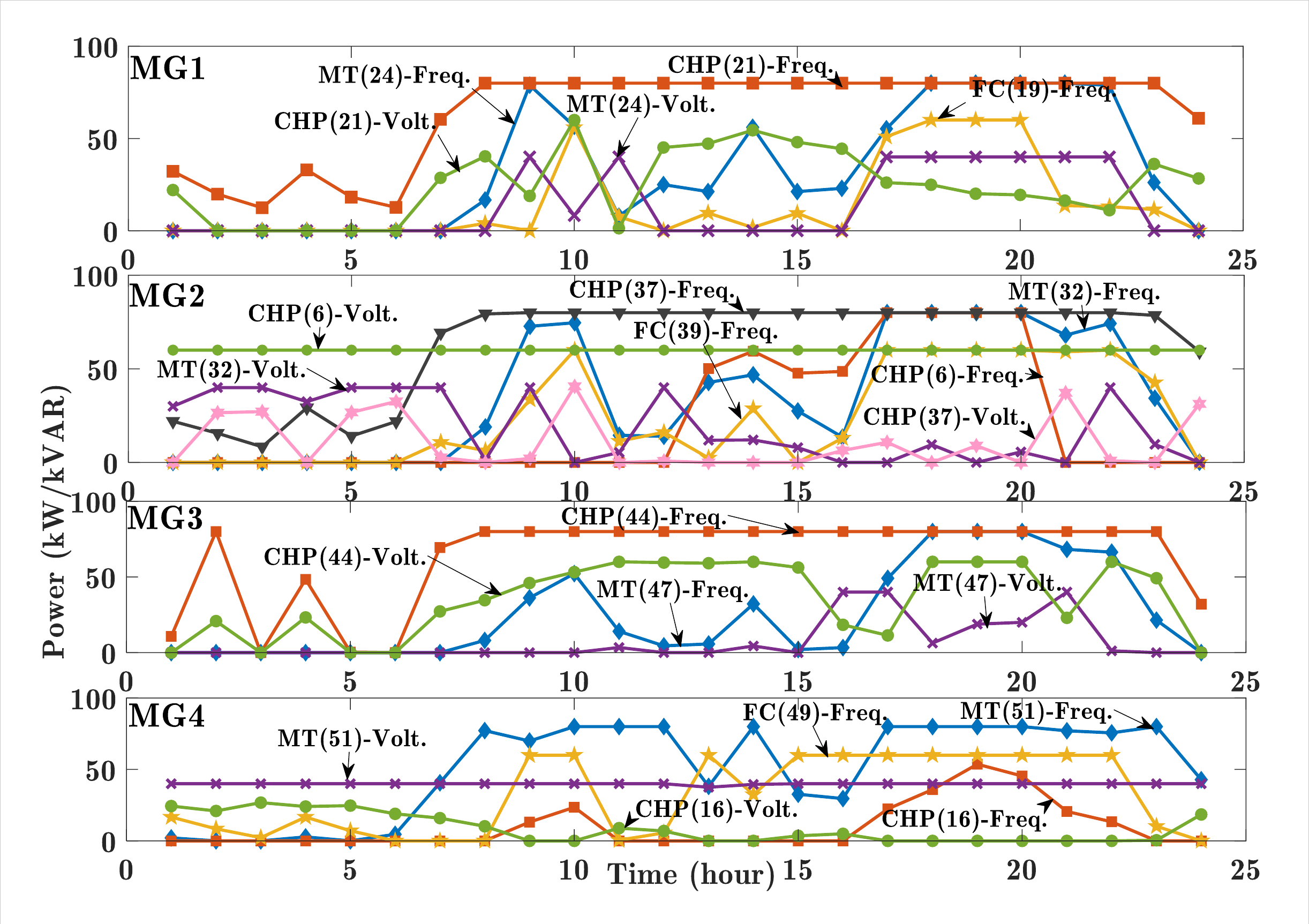}
	\caption{\footnotesize Contribution of droop control on real and reactive power}
	\label{fig10} 
\end{figure}

According to Fig. \ref{fig08} and Fig. \ref{fig09}, CHP and MT units have a significant contribution in providing the real and reactive loads in each MG. Since dispatchable DERs are equipped with droop controllers, MT, FC, and CHP generators contribute to frequency-based droop control, in which frequency variations can cause changes in generation level of DERs, while MT and CHP generators are used for voltage-based droop control to adjust the reactive power in a predetermined range. In Fig. \ref{fig10}, the contribution of droop control of DERs in each MG is illustrated.

\begin{figure}[H]
	\centering
	\includegraphics[width=0.5\textwidth]{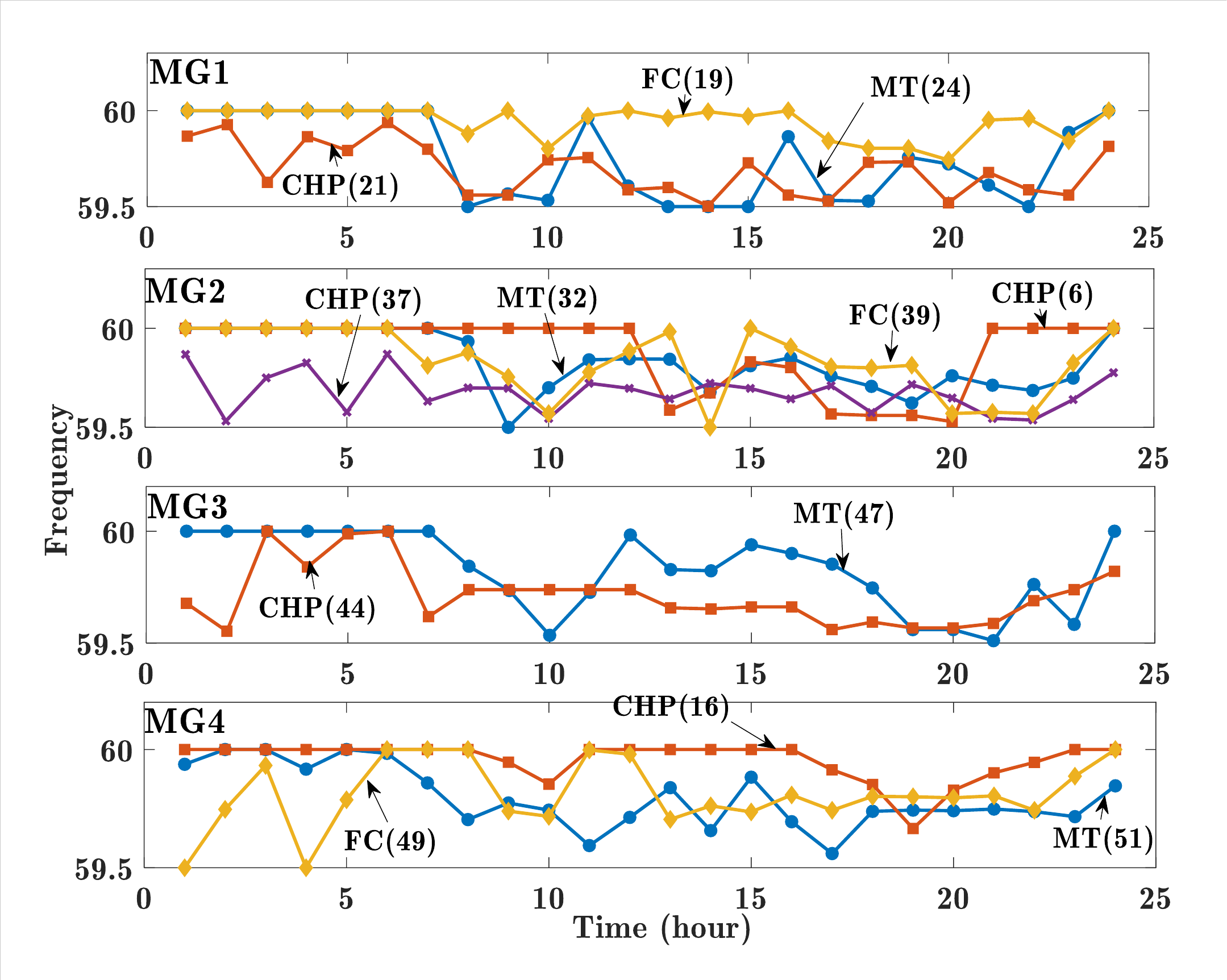}
	\caption{\footnotesize Frequency fluctuation due to droop control performance in each DER}
	\label{fig13} 
\end{figure}

In Fig. \ref{fig10}, generally, the contribution of frequency-based droop control in generating real power for a DER is higher than the contribution of voltage-based droop control in reactive power generation for the same DER. The reason is that the maximum real power generation capacity is more than the maximum capacity of DERs in reactive power generation (see Table \ref{table02}). Precisely, the real and reactive power balances are deciding factors of the contribution of droop controllers in the power generation of DERs. The frequency fluctuation of each DER is caused by frequency-based droop control is shown in Fig. \ref{fig13}.

\begin{figure}[H]
	\centering
	\includegraphics[width=0.5\textwidth]{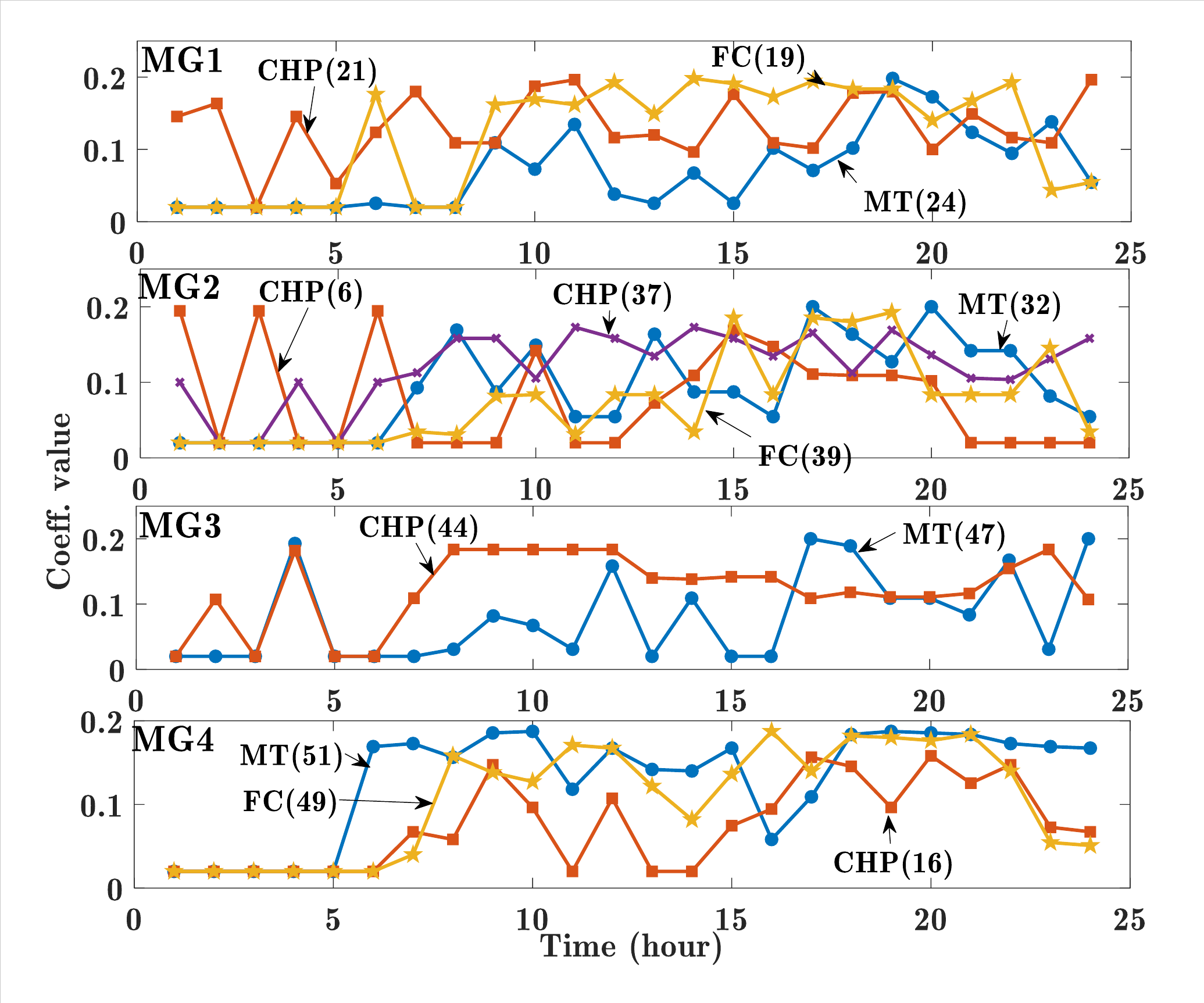}
	\caption{\footnotesize Optimized values for real power droop control coefficients obtained by DA-SLR method}
	\label{fig11} 
\end{figure}

\begin{figure}[H]
	\centering
	\includegraphics[width=0.5\textwidth]{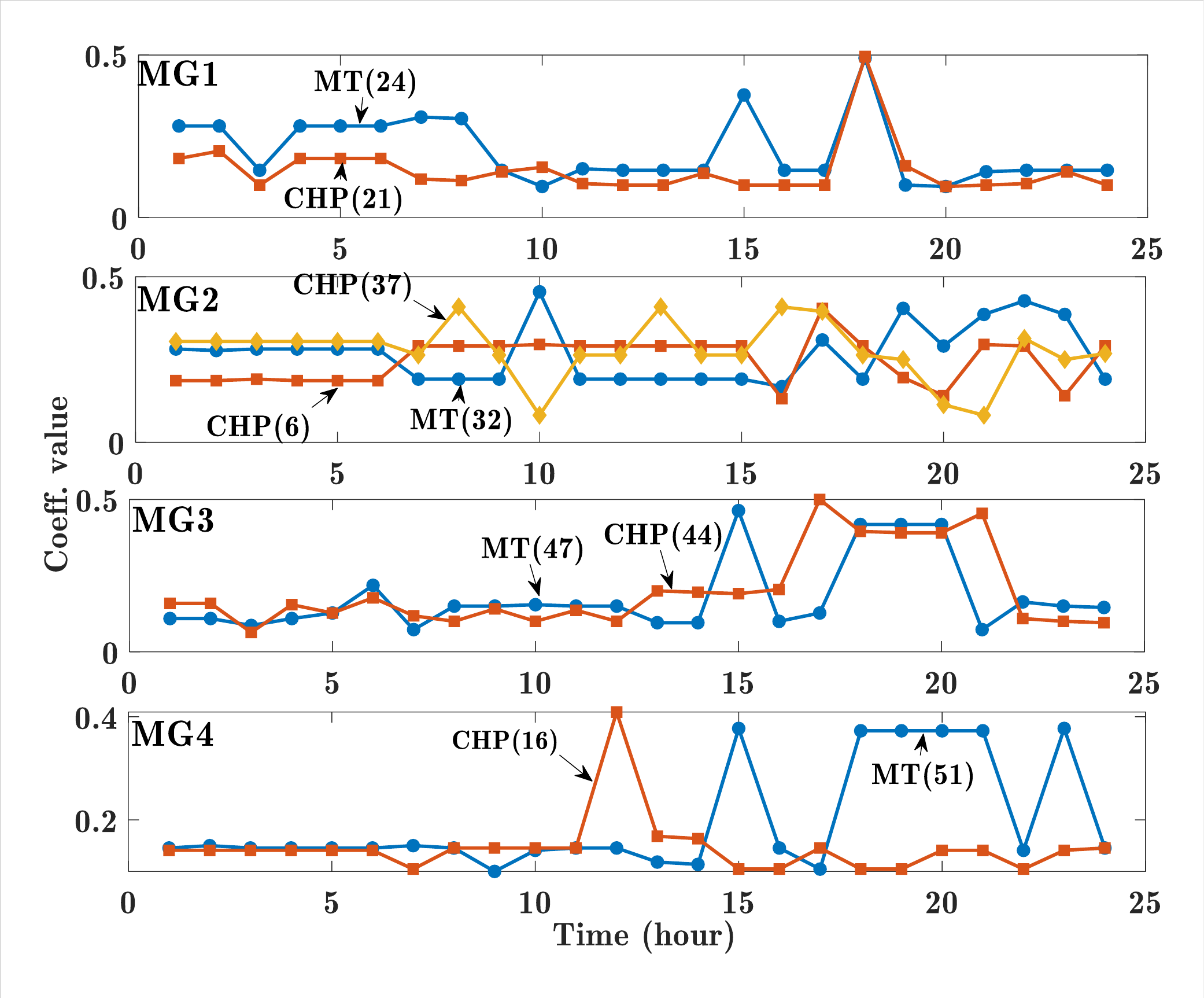}
	\caption{\footnotesize Optimized values for reactive power droop control coefficients obtained by DA-SLR method}
	\label{fig12} 
\end{figure}

Comparing Fig. \ref{fig10} and Fig. \ref{fig13} shows that the DER decreases its frequency to enhance its droop control contribution in real power generation and vice versa. Take the CHP unit in MG1 as an example. As shown in Fig. \ref{fig10}, CHP contribution using frequency-based droop control is in the maximum level after hour $7$, which results in the lowest frequency level comparing to previous hours according to Fig. \ref{fig13}. Also, as the frequency and voltage droop coefficients are considered discrete variables, the optimized values for droop coefficients is shown in Figs.~\ref{fig11} and~\ref{fig12}.

A droop controller prefers to adjust the frequency and voltage droop coefficients in the higher or lower levels or step to keep the power balance guaranteed. Comparing the Fig. \ref{fig10} and Fig. \ref{fig11} for frequency-based droop control shows that when the real power generation of a DER increases the frequency-based droop control coefficient increases compared to the hours that the generated real power is lower. Similarly, the reactive power generation level of a DER in Fig. \ref{fig09} has a direct relationship with the contribution of voltage-based droop control in Fig. \ref{fig10} thereby the droop control coefficient dealing with reactive power will be changed to guarantee the reactive power balance. Take MT(24) of MG1 as an example, where a direct relationship between the reactive power generation and corresponding droop control coefficient is observable.

Moreover, to show the impact of droop controller contribution in this study, the applied technical limit on droop control coefficients is enhanced from 20 percent to 30 percent of the total generation of each DER in the next scenario. In Table \ref{table04}, this impact is demonstrated.
{\rowcolors{2}{gray!10}{lightgray!2}
\begin{table}[h]
    \centering
    \caption{Impact of droop control contribution increment on the operation cost using DA-SLR}
    \begin{tabular}{|p{1.6cm}||p{0.9cm}||p{0.9cm}||p{0.9cm}||p{0.9cm}||p{0.9cm}|}
    \rowcolor{gray!60}  & \multicolumn{5}{|c|}{Entity} \\
     & MG1 & MG2 & MG3 & MG4 & Total \\
    \hline\hline
    \small{Objective function (\$)} &  $1711$ & $4063$  &  $2563$ &  $2410$ &  $10747$\\
    \small{Reduction (\%)}  &  $19.2$ & $22.3$  &  $18.6$ &  $21.3$ &  $20.7$\\
    \hline
    \end{tabular}
    \label{table04}
\end{table}
}

In Table \ref{table04}, the importance of droop control coefficients and consequently the optimized values are shown. In the optimization process, changing droop control coefficients limits can have a profound impact on the operation cost of microgrids.

\subsection{Test results on a 123-bus networked microgrids system}\label{05:02}
\begin{figure}[H]
	\centering
	\includegraphics[width=0.48\textwidth]{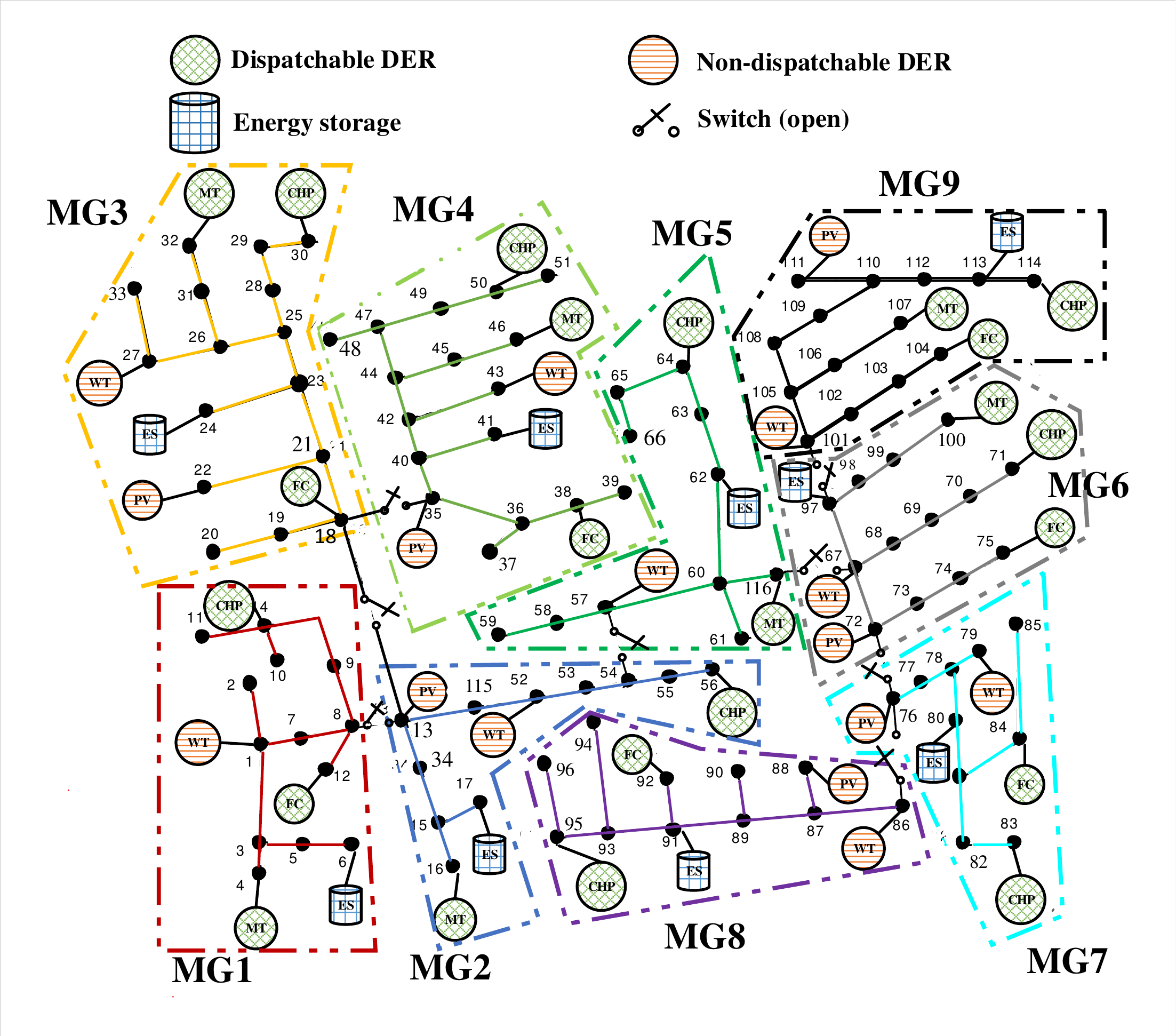}
	\caption{\small Modified IEEE 123-bus distribution system with nine MGs}
	\label{fig05} 
\end{figure}

\begin{figure}[H]
	\centering
	\includegraphics[width=0.5\textwidth]{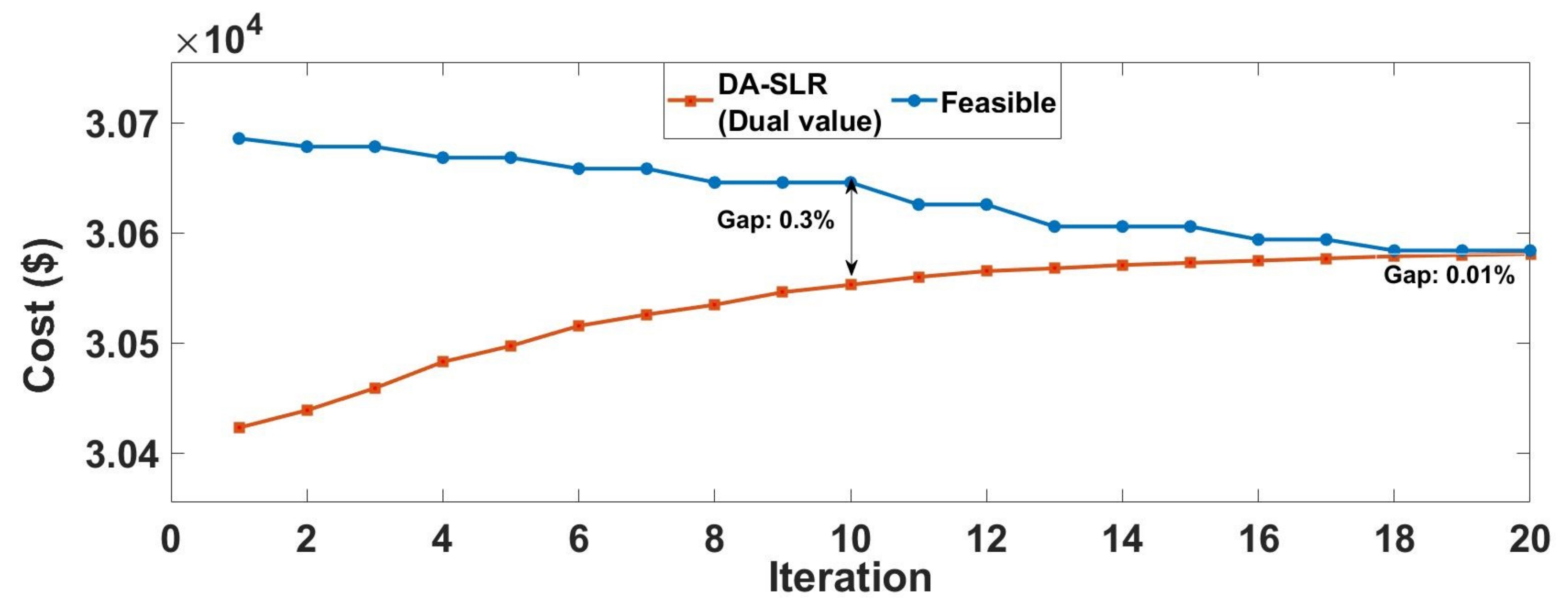}
	\caption{\footnotesize Tracking the feasible solution using SLR and DA-SLR in modified IEEE 123-bus system}
	\label{fig14} 
\end{figure}

To demonstrate the scalability of DA-SLR, a larger network is considered. Fig. \ref{fig05} shows the single line diagram of a 123-bus networked microgrids system with nine interconnected MGs.
The  results obtained by the designed DA-SLR method are analyzed upon approaching the feasible cost (see Fig. \ref{fig14}). The gap reaches 0.3\% and 0.01\% after 10 and 20 iterations respectively. 
In Table \ref{table05}, the operation cost of MGs is analyzed using three methods, DA-SLR, sequential SLR, and ADMM.


{\rowcolors{2}{gray!10}{lightgray!2}
\begin{table}[h]
    \centering
    \caption{Impact of droop control contribution increment on the operation cost using DA-SLR}
    \begin{tabular}{|p{1.6cm}||p{0.9cm}||p{0.9cm}||p{0.9cm}||p{0.9cm}||p{0.9cm}|}
    \rowcolor{gray!60}  Method & \multicolumn{5}{|c|}{Objective function (\$)} \\
     & MG1 & MG2 & MG3 & MG4 & MG5 \\
    \hline\hline
    \small{DA-SLR} &  $4109$ & $950$  &  $4925$ &  $677$ &  $3677$\\
    \small{ADMM}  &  $7289$ & $-$  &  $5537$ &  $5799$ &  $-$\\
    \rowcolor{gray!60}  & \multicolumn{5}{|c|}{Objective function (\$)} \\
     & MG6 & MG7 & MG8 & MG9 & Total \\
     \hline
    \small{DA-SLR} &  $4452$ & $2668$  &  $6357$ &  $2744$ &  $30559$\\
    \small{ADMM}  &  $-$ & $-$  &  $7001$ &  $9555$ &  $-$\\
    \hline
    \end{tabular}
    \label{table05}
\end{table}
}

According to the results of Table \ref{table05}, the DA-SLR has a better performance of optimization compared to sequential SLR and ADMM. It is seen that the ADMM method suffers from divergence problems in solving some subproblems.

The DA-SLR and ADMM spend 35 and 210 seconds per iteration to solve MGs' subproblems, respectively. This comparison shows the definite advantage of DA-SLR as compared to the ADMM.  
\section{Conclusion}\label{Section:06}
In this paper, the distributed and asynchronous surrogate Lagrangian relaxation method is used to coordinate networked microgrids to schedule the islanded MGs in a distributed manner. The new method efficiently handles binary decisions and excellent performance, as well as scalability, is demonstrated. The method paves the way to facilitate software-defined networking 
in enabling efficient coordination of distributed entities asynchronously. It was shown that the DA-SLR method can achieve acceptable convergence in fewer iterations and guarantee a minimum amount of operation cost. The classical method ADMM, in contrast, could not efficiently manage a distributed problem when the number of microgrids or subproblems are growing. Moreover, discrete variables have no impact on the convergence of the DA-SLR method, while classical methods face difficulties in solving problems with discrete decision variables. The next step is to discuss the potential of the DA-SLR method on stability issues of operational optimization of programmable microgrids.


\bibliographystyle{IEEEtran}
\bibliography{paperBibliography}

\begin{thebibliography}{10}
\providecommand{\url}[1]{#1}
\csname url@samestyle\endcsname
\providecommand{\newblock}{\relax}
\providecommand{\bibinfo}[2]{#2}
\providecommand{\BIBentrySTDinterwordspacing}{\spaceskip=0pt\relax}
\providecommand{\BIBentryALTinterwordstretchfactor}{4}
\providecommand{\BIBentryALTinterwordspacing}{\spaceskip=\fontdimen2\font plus
\BIBentryALTinterwordstretchfactor\fontdimen3\font minus
  \fontdimen4\font\relax}
\providecommand{\BIBforeignlanguage}[2]{{%
\expandafter\ifx\csname l@#1\endcsname\relax
\typeout{** WARNING: IEEEtran.bst: No hyphenation pattern has been}%
\typeout{** loaded for the language `#1'. Using the pattern for}%
\typeout{** the default language instead.}%
\else
\language=\csname l@#1\endcsname
\fi
#2}}
\providecommand{\BIBdecl}{\relax}
\BIBdecl

\bibitem{NSFproposal}
P.~Zhang, P.~Luh, C.~Atkinson-Palombo, B.~Li, and A.~Herzberg, \emph{SCC:
  Empowering Smart and Connected Communities through Programmable Community
  Microgrids}, Proposal for Grant ECCS-1831811/2018492, National Science
  Foundation, Sept. 2018.

\bibitem{6994333}
D.~{Kreutz}, F.~M.~V. {Ramos}, P.~E. {Veríssimo}, C.~E. {Rothenberg},
  S.~{Azodolmolky}, and S.~{Uhlig}, ``Software-defined networking: A
  comprehensive survey,'' \emph{Proceedings of the IEEE}, vol. 103, no.~1, pp.
  14--76, 2015.

\bibitem{zhang2021networked}
P.~Zhang, \emph{Networked Microgrids}, Cambridge University Press, 2021.

\bibitem{zhang2019enabling}
P.~Zhang, B.~Wang, P.~B. Luh, L.~Ren, and Y.~Qin, ``Enabling resilient
  microgrid through ultra-fast programmable network,'' {US Patent No.
  10,505,853, Date of Patent: Dec. 10, 2019}.

\bibitem{9099877}
L.~{Wang}, Y.~{Qin}, Z.~{Tang}, and P.~{Zhang}, ``Software-defined microgrid
  control: The genesis of decoupled cyber-physical microgrids,'' \emph{IEEE
  Open Access Journal of Power and Energy}, vol.~7, pp. 173--182, 2020.

\bibitem{4488380}
L.~{Balasevicius}, A.~{Kunickaite}, and V.~S. {Janusevicius}, ``Discrete-time
  pid controller design in programmable logical controllers,'' in \emph{2007
  4th IEEE Workshop on Intelligent Data Acquisition and Advanced Computing
  Systems: Technology and Applications}, 2007, pp. 86--90.

\bibitem{7870716}
W.~{Gu}, G.~{Lou}, W.~{Tan}, and X.~{Yuan}, ``A nonlinear state estimator-based
  decentralized secondary voltage control scheme for autonomous microgrids,''
  \emph{IEEE Transactions on Power Systems}, vol.~32, no.~6, pp. 4794--4804,
  2017.

\bibitem{7592418}
X.~{Lu}, X.~{Yu}, J.~{Lai}, Y.~{Wang}, and J.~M. {Guerrero}, ``A novel
  distributed secondary coordination control approach for islanded
  microgrids,'' \emph{IEEE Transactions on Smart Grid}, vol.~9, no.~4, pp.
  2726--2740, 2018.

\bibitem{6980137}
W.~{Shi}, X.~{Xie}, C.~{Chu}, and R.~{Gadh}, ``Distributed optimal energy
  management in microgrids,'' \emph{IEEE Transactions on Smart Grid}, vol.~6,
  no.~3, pp. 1137--1146, 2015.

\bibitem{GEORGES1994155}
D.~Georges, ``Optimal unit commitment in simulations of hydrothermal power
  systems: an augmented lagrangian approach,'' \emph{Simulation Practice and
  Theory}, vol.~1, no.~4, pp. 155 -- 172, 1994.

\bibitem{HADIAMINI2018137}
M.~{Hadi Amini}, S.~Bahrami, F.~Kamyab, S.~Mishra, R.~Jaddivada, K.~Boroojeni,
  P.~Weng, and Y.~Xu, ``Chapter 6 - decomposition methods for distributed
  optimal power flow: Panorama and case studies of the dc model,'' in
  \emph{Classical and Recent Aspects of Power System Optimization}.\hskip 1em
  plus 0.5em minus 0.4em\relax Academic Press, 2018, pp. 137 -- 155.

\bibitem{5960802}
T.~{Erseghe}, D.~{Zennaro}, E.~{Dall'Anese}, and L.~{Vangelista}, ``Fast
  consensus by the alternating direction multipliers method,'' \emph{IEEE
  Transactions on Signal Processing}, vol.~59, no.~11, pp. 5523--5537, 2011.

\bibitem{8682117}
P.~{Ramanan}, M.~{Yildirim}, E.~{Chow}, and N.~{Gebraeel}, ``An asynchronous,
  decentralized solution framework for the large scale unit commitment
  problem,'' \emph{IEEE Transactions on Power Systems}, vol.~34, no.~5, pp.
  3677--3686, 2019.

\bibitem{9112667}
M.~A. {Bragin}, B.~{Yan}, and P.~B. {Luh}, ``Distributed and asynchronous
  coordination of a mixed-integer linear system via surrogate lagrangian
  relaxation,'' \emph{IEEE Transactions on Automation Science and Engineering},
  pp. 1--15, 2020.

\bibitem{9007663}
W.~{Wan}, M.~{Bragin}, B.~{Yan}, Y.~{Qin}, J.~{Philhower}, P.~{Zhang}, and
  P.~{Luh}, ``Distributed and asynchronous active fault management for
  networked microgrids,'' \emph{IEEE Transactions on Power Systems}, pp. 1--1,
  2020.

\bibitem{4118327}
N.~{Pogaku}, M.~{Prodanovic}, and T.~C. {Green}, ``Modeling, analysis and
  testing of autonomous operation of an inverter-based microgrid,'' \emph{IEEE
  Transactions on Power Electronics}, vol.~22, no.~2, pp. 613--625, 2007.

\bibitem{6152194}
H.~{Yeh}, D.~F. {Gayme}, and S.~H. {Low}, ``Adaptive var control for
  distribution circuits with photovoltaic generators,'' \emph{IEEE Transactions
  on Power Systems}, vol.~27, no.~3, pp. 1656--1663, 2012.

\bibitem{6497085}
S.~{Tan}, J.~{Xu}, and S.~K. {Panda}, ``Optimization of distribution network
  incorporating distributed generators: An integrated approach,'' \emph{IEEE
  Transactions on Power Systems}, vol.~28, no.~3, pp. 2421--2432, 2013.

\bibitem{6895183}
Z.~{Wang}, H.~{Chen}, J.~{Wang}, and M.~{Begovic}, ``Inverter-less hybrid
  voltage/var control for distribution circuits with photovoltaic generators,''
  \emph{IEEE Transactions on Smart Grid}, vol.~5, no.~6, pp. 2718--2728, 2014.

\bibitem{Sherali2013}
H.~D. Sherali and W.~P. Adams, \emph{Reformulation--Linearization Techniques
  for Discrete Optimization Problems}.\hskip 1em plus 0.5em minus 0.4em\relax
  New York, NY: Springer New York, 2013, pp. 2849--2896.

\bibitem{25627}
M.~E. {Baran} and F.~F. {Wu}, ``Network reconfiguration in distribution systems
  for loss reduction and load balancing,'' \emph{IEEE Transactions on Power
  Delivery}, vol.~4, no.~2, pp. 1401--1407, 1989.

\bibitem{119237}
W.~H. {Kersting}, ``Radial distribution test feeders,'' \emph{IEEE Transactions
  on Power Systems}, vol.~6, no.~3, pp. 975--985, 1991.

\end{thebibliography}


\end{document}